\documentclass[twocolumn]{aastex62}
\usepackage{amsmath}
\usepackage{float}

\graphicspath{{./}{figures/}}

\shorttitle{A Real-Time Search for Interstellar Impacts on the Moon}

\begin{document}

\title{A Real-Time Search for Interstellar Impacts on the Moon}

\email{amir.siraj@cfa.harvard.edu, aloeb@cfa.harvard.edu}

\author{Amir Siraj}
\affil{Department of Astronomy, Harvard University, 60 Garden Street, Cambridge, MA 02138, USA}

\author{Abraham Loeb}
\affiliation{Department of Astronomy, Harvard University, 60 Garden Street, Cambridge, MA 02138, USA}




\begin{abstract}
The discovery of `Oumuamua and CNEOS 2014-01-08 allowed for a calibration of the impact rate of interstellar objects. We propose a new telescope in lunar orbit to study in real-time interstellar meteoroid impacts and to serve as a laboratory for hypervelocity collisions. We show that a telescope with diameter $D_a \gtrsim 2 \mathrm{\; m}$ would be capable of detecting $\gtrsim 1$ interstellar meteoroid impacts (among hundreds of Solar System meteoroid impacts) per year. For each meteoroid, measurements of the reflected sunlight and shadow, as well as the impact's optical flash and crater, would allow for the determination of the 3D velocity, mass, density, and composition, as well as the radiative efficiency.

\end{abstract}



\section{Introduction}

`Oumuamua was the first interstellar object detected in the Solar System \citep{Meech2017, Micheli2018}. Several follow-up studies of `Oumuamua were conducted to better understand its origin and composition \citep{Bannister2017, Gaidos2017, Jewitt2017, Mamajek2017, Ye2017, Bolin2017, Fitzsimmons2018, Trilling2018, Bialy2018, Hoang2018, Siraj2019a, Siraj2019b, Seligman2019}. `Oumuamua's size was estimated to be 20m - 200m, based on Spitzer Space Telescope constraints on its infrared emission given its expected surface temperature based on its orbit \citep{Trilling2018}.

CNEOS 2014-01-08 \citep{Siraj2019c} is tentatively the first interstellar meteor discovered larger than dust \citep{Baggaley1993, Hajdukova1994, Taylor1996, Baggaley2000, Mathews1998, Meisel2002a, Meisel2002b, Weryk2004, Afanasiev2007, Musci2012, Engelhardt2017, Hajdukova2018}, allowing for a calibration of the local flux of such objects. Whereas on Earth small meteoroids burn up high in the atmosphere, on the Moon they directly impact the surface, generating bright flashes of light and craters \citep{Rubio2000, Burchell2010, Goel2015, Ortiz2015, Ortiz2016, Avdellidou2019}.

In this paper, we propose a mission dedicated to the discovery and characterization of interstellar impacts on the Moon. We show that a lunar satellite equipped with a telescope to monitor such events in real time could allow for precise velocity, mass, and density determination of interstellar objects. Immediate follow-up spectroscopy as well as exploration of the resulting craters could reveal detailed composition of the impactors. Such data would calibrate population parameters for interstellar objects crucial for constraining theories of planetary formation and for assessing the habitability of exoplanets \citep{Duncan1987, Charnoz2003, Veras2011, Veras2014, Pfalzner2015, Do2018, Raymond2018,  Hands2019, Pfalzner2019, Siraj2019e}, as well as the search for the building blocks of extraterrestrial life \citep{Lingam2019}. Such a satellite could also serve as a laboratory for hypervelocity impacts \citep{Rubio2000, Drolshagen2001, Burchell2010, Ortiz2015, Ortiz2016, Goel2015, Sachse2015, Avdellidou2019}, constraining physical processes difficult to replicate on Earth.

The outline of the paper is as follows. In Section~\ref{sec:methods}, we explore the detectable impact rate (\ref{ssec:rate}), as well as the relevant processes for 3D velocity determination (\ref{ssec:velocity}), cratering (\ref{ssec:cratering}), and optical flashes (\ref{ssec:flashes}). In Section~\ref{sec:results} we report our results, and in Section~\ref{sec:discussion} we discuss the implications of these results. 

\section{Method}
\label{sec:methods}

\subsection{Detectable Impact Rate}
\label{ssec:rate}
Throughout this paper we consider a satellite orbiting at altitude $h = 100 \; \mathrm{km}$ above the lunar surface, equipped with a telescope with aperture diameter $D_a$. The corresponding diffraction-limited spatial resolution $\Delta l$ at visible wavelengths, $\lambda \sim 500 \; \mathrm{nm}$, on the lunar surface at a distance $z$ away from the satellite is,

\begin{equation}
    \label{eq:visible}
    \Delta l \;\approx \;6 \times 10^{-2} \left( \frac{D_a}{1 \mathrm{\; m}}\right)^{-1} \left( \frac{z}{1 \mathrm{\; km}}\right) \; \mathrm{cm} \; \;.
\end{equation}

Given that CNEOS 2014-01-08 implies an Earth impact rate of $\sim 0.1 \mathrm{\; yr^{-1}}$ for interstellar meteoroids of size $d \sim 1 \; \mathrm{m}$ \citep{Siraj2019c} and that the power law exponent for the cumulative size distributions of interstellar meteoroids is $\sim 3.4$ \citep{Siraj2019d, Musci2012, Landgraf2000}, we estimate the impact rate of centimeter-scale meteoroids with diameter $d$ on the moon to be,

\begin{equation}
    \dot{n} \; \sim  \; 4 \times 10^{-3} \left( \frac{d}{1 \mathrm{\; cm}}\right)^{-3.4} \; \mathrm{km^{-2} \; yr^{-1}} \; \;.
\end{equation}

Substituting the dependence of the spatial resolution on aperture size and distance, we derive the impact rate,

\begin{equation}
    \label{eq:ndot}
    \dot{n} \; \sim  \; 60 \left( \frac{D_a}{1 \mathrm{\; m}}\right)^{3.4} \left( \frac{z}{1 \mathrm{\; km}}\right)^{-3.4} \; \mathrm{km^{-2} \; yr^{-1}} \; \;.
\end{equation}

\begin{figure}
  \centering
  \includegraphics[width=.8\linewidth]{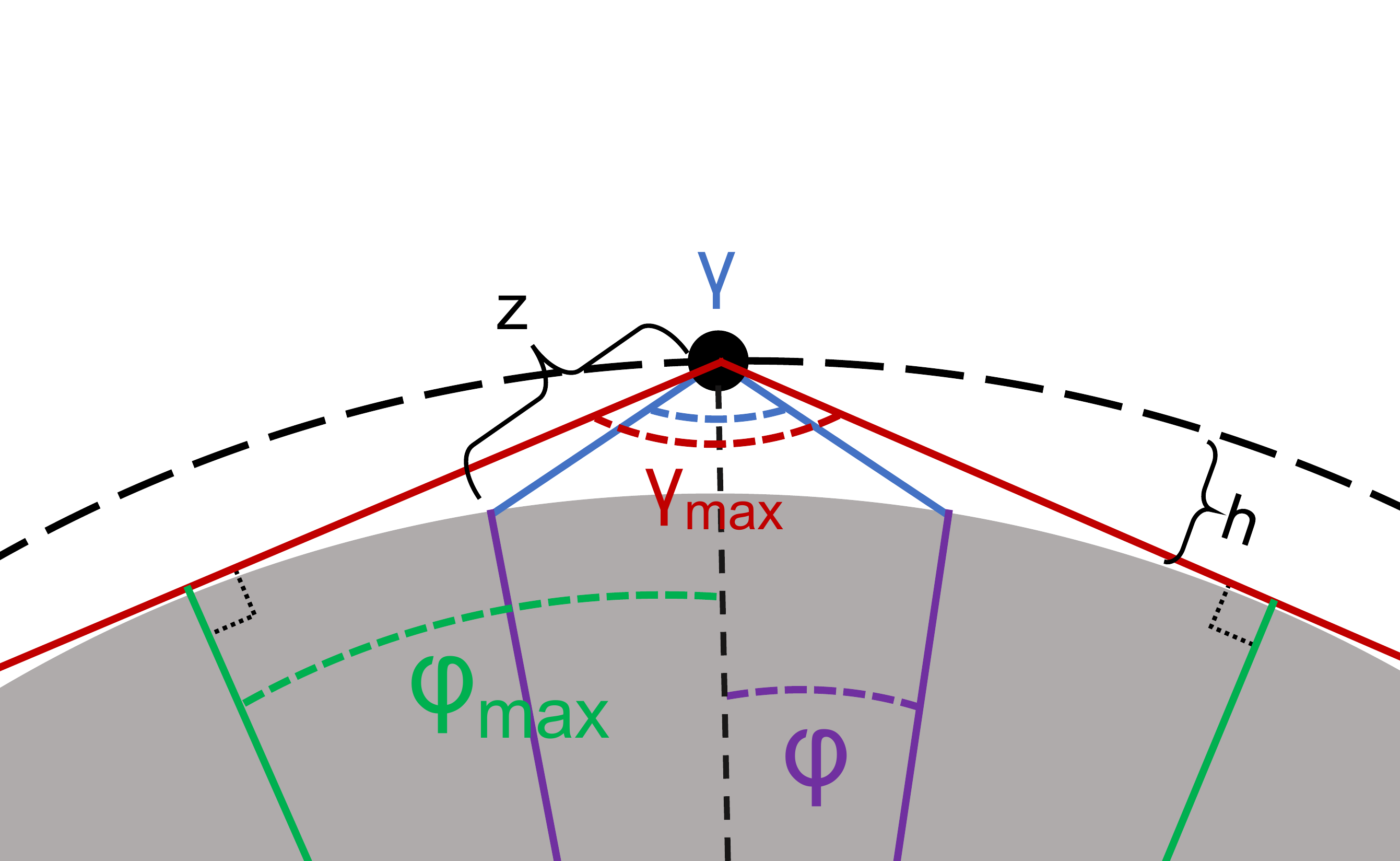}
    \caption{The telescope's orbit is indicated by the dashed line at an altitude $h$, with $z$ being its distance from a point of interest on the lunar surface at a lunar latitude separation $\phi$, requiring the telescope to be capable of viewing out to an angle of $\gamma$ in any direction. The angles $\phi_{max}$ and $\gamma_{max}$ are associated with the furthest visible point on the moon from the telescope.}
    \label{fig:diagram}
\end{figure}

We adopt the setup illustrated in Figure~\ref{fig:diagram}, in which the field-of-view (FOV) of the satellite is represented by the solid angle $2\gamma$ and the visible fraction of the lunar surface is represented by the angle $2\phi = 2\tan^{-1}\left({\frac{R\sin{\gamma/2}}{R + h - R\cos{\gamma/2}}}\right)$. For altitude $h = 100 \; \mathrm{km}$, $\gamma_{max} \approx 0.33 \mathrm{\; rad}$ and $\phi_{max} \approx 1.24 \mathrm{\; rad}$,\footnote{The angles associated with the furthest visible point on the moon from the telescope, as indicated in Figure \ref{fig:diagram}.} the total rate of detectable impacts as a function of $\gamma$ is,

\begin{equation}
    \dot{N} \sim \int_0^{2\pi} \int_0^{\tan^{-1}\left({\frac{R\sin{\gamma/2}}{R + h - R\cos{\gamma/2}}}\right)} \dot{n}\; R^2 \sin{\phi} \; \mathrm{d\phi \; d\theta} \; \; ,
\end{equation}
where the radius of the Moon $R \approx 1700 \mathrm{\; km}$, and,

\begin{equation}
    z = \sqrt{R\sin(\phi)^2 + (R + h - R\cos(\phi))^2} \; \; .
\end{equation}

\subsection{Velocity Determination}
\label{ssec:velocity}
The speed and direction of the meteoroid are computed by tracking the changes in position as a function of time for both the meteoroid and its shadow on the lunar surface prior to impact. While this technique requires the Sun to be shining on the area of interest, it should reliably produce provide the 3D velocity of the meteoroid, which is usually impossible to measure with a single telescope. Since interstellar objects travel at several tens of $\mathrm{\; km \; s^{-1}}$, and the telescope is located at an altitude of $h = \mathrm{100 \; km}$, a typical crossing time of the object's size is a few seconds. The approach would be to subtract frames and only search for time-dependent features in the image (reducing the amount of data stored). The length of the streaks on the subtracted frames will indicate the meteoroid's velocity and the width of the streaks will indicate its size.

Although Eq.~(\ref{eq:visible}) guarantees that the spatial resolution of the telescope is sufficient to detect a meteoroid prior to impact, we must check that the signal from the meteoroid is sufficiently strong. The following order-of-magnitude calculation applies to detecting the shadow of the passing meteoroid.

The Solar flux is $\sim 10^{18} \mathrm{\; photons \; cm^{-2} \; s^{-1}}$, and since the Moon's albedo is $\sim 10\%$, we take the flux from the Moon's surface to be $10^{17} \mathrm{\; photons \; s^{-1}}$. The rate of photons received by a telescope with a collecting area of $\sim 10^4 \mathrm{\;cm^2}$ at a distance of $10^7 \mathrm{\;cm}$ from an area $A$ is $10^6 (A/\mathrm{cm^2}) \mathrm{\; photons \; s^{-1}}$. A meteoroid with speed $\sim 50 \mathrm{\; km \; s^{-1}}$ at an angle of $\sim 45^{\circ}$ has an apparent speed of $\sim 35 \mathrm{\; km \; s^{-1}}$, and its shadow blocks a region of area $A$ for $3 \times 10^{-7} \; \sqrt{(A/\mathrm{cm^2})} \mathrm{\; s}$. Therefore the number of photons blocked by the meteoroid's shadow for each region of area $A$ is $\sim 0.3 \; \left( \frac{A}{1 \; \mathrm{cm^2}} \right)^{3/2} \; \mathrm{photons}$.

For example, a meteoroid of diameter $d \sim 4 \mathrm{\; cm}$ would block $\sim 10 \; \mathrm{photons}$ over each region of area $A \sim 10 \mathrm{\; cm^2}$ for a period of $\sim 10^{-6} \mathrm{\; s}$. Assuming a read-out noise\footnote{Comparable to the read-out rates of the upcoming Large Synoptic Survey Telescope and Hubble Space Telescope's Advanced Camera for Surveys.} of $\sim 5\; e^-$, a frame rate of $\sim 10^6 \mathrm{\; s^{-1}}$ would allow for $\mathrm{S/N} \sim 2$ for each area of blocked light. Similarly, as long as an object's albedo is different from that of the lunar surface, one could directly detect its reflected sunlight. Surface color and albedo of interstellar objects are likely to be different from the surface of the Moon as interstellar objects are exposed to interstellar space and not protected by the Solar wind. Furthermore, the curvature of the interstellar objects' surfaces will cause them to reflect light differently than the lunar surface.

\subsection{Cratering}
\label{ssec:cratering}
After a meteoroid detection, the frame rate can be automatically slowed down to an appropriate pace to discern the diameter of the resulting crater.

Crater diameter $D_c$ as a function of regolith density $\rho_r$, meteoroid density $\rho_m$, meteoroid energy $E$, and impact angle $\theta$ for meteoroids smaller than $\sim 100 \mathrm{\; m}$ is estimated by \cite{Gault1974} and \cite{Melosh1989}, and utilized by \cite{Suggs2014} and \cite{Ortiz2015} for lunar impacts, as,

\begin{equation}
\begin{aligned}
    D_c \approx 3.8 \mathrm{\; m} 
    & \left(\frac{\rho_m}{\mathrm{1 \; g \; cm^{-3}}}\right)^{1/6} \left(\frac{\rho_r}{\mathrm{1 \; g \; cm^{-3}}}\right)^{-1/2}
    \\
    & \left(\frac{E}{\mathrm{10^{15} \; ergs}}\right)^{0.29} \sin(\theta)^{1/3} \; \;.
\end{aligned}
\end{equation}

We can solve the above expression to find the meteoroid density $\rho_m$ as a function of the other parameters,

\begin{equation}
\begin{aligned}
    \rho_m \approx 1 \mathrm{\; g \; cm^{-3}}  & 
     \left(\frac{d}{\mathrm{5.4 \; cm}}\right)^{-5.22} \left(\frac{v}{\mathrm{50 \; km \; s^{-1}}}\right)^{-3.48} 
    \\
    & \left(\frac{D_c}{\mathrm{3.8 \; m}}\right)^{6} \left(\frac{\rho_r}{\mathrm{1 \; g \; cm^{-3}}}\right)^{3}
    \sin(\theta)^{-2} \; \; ,
\end{aligned}
\end{equation}
where $v$ is the impact speed.

\subsection{Optical Flashes}
\label{ssec:flashes}
Meteoroid impacts on the Moon give rise to optical flashes that last for tens of milliseconds, with energy equal to $\zeta E$, where $E$ is the kinetic energy of the meteoroid and $\zeta$ is the luminous efficiency \citep{Avdellidou2019}. The luminous efficiency of hypervelocity impacts is an active field of research  \citep{Rubio2000, Goel2015, Ortiz2015, Ortiz2016, Avdellidou2019}. Since the kinetic energy is constrained by the crater size, we can derive the luminous efficiency of the impact by comparing the computed kinetic energy and the observed energy of the optical flash.

A standard scientific camera with a frame rate of $\sim 10^3 \mathrm{\; s^{-1}}$ can easily measure the optical flash from the satellite's orbit. Assuming $\zeta \sim 10^{-3}$ \citep{Ortiz2015}, $\rho_m \sim 2 \mathrm{\; g \; cm^{-3}}$, $d \sim 4 \; \mathrm{cm}$, and emitting area $\sim 10^4 \mathrm{\; cm^2}$, the optical flash is several $\sim 10^4$ times brighter than reflected sunlight from the lunar surface. 

Spectroscopy of resulting gas plumes potentially reveal the compositions of meteoroids. Isotope ratios can be used to determine if the origin is from the Solar System \citep{Lingam2019}.

\section{Results}
\label{sec:results}

\begin{figure}
  \centering
  \includegraphics[width=.8\linewidth]{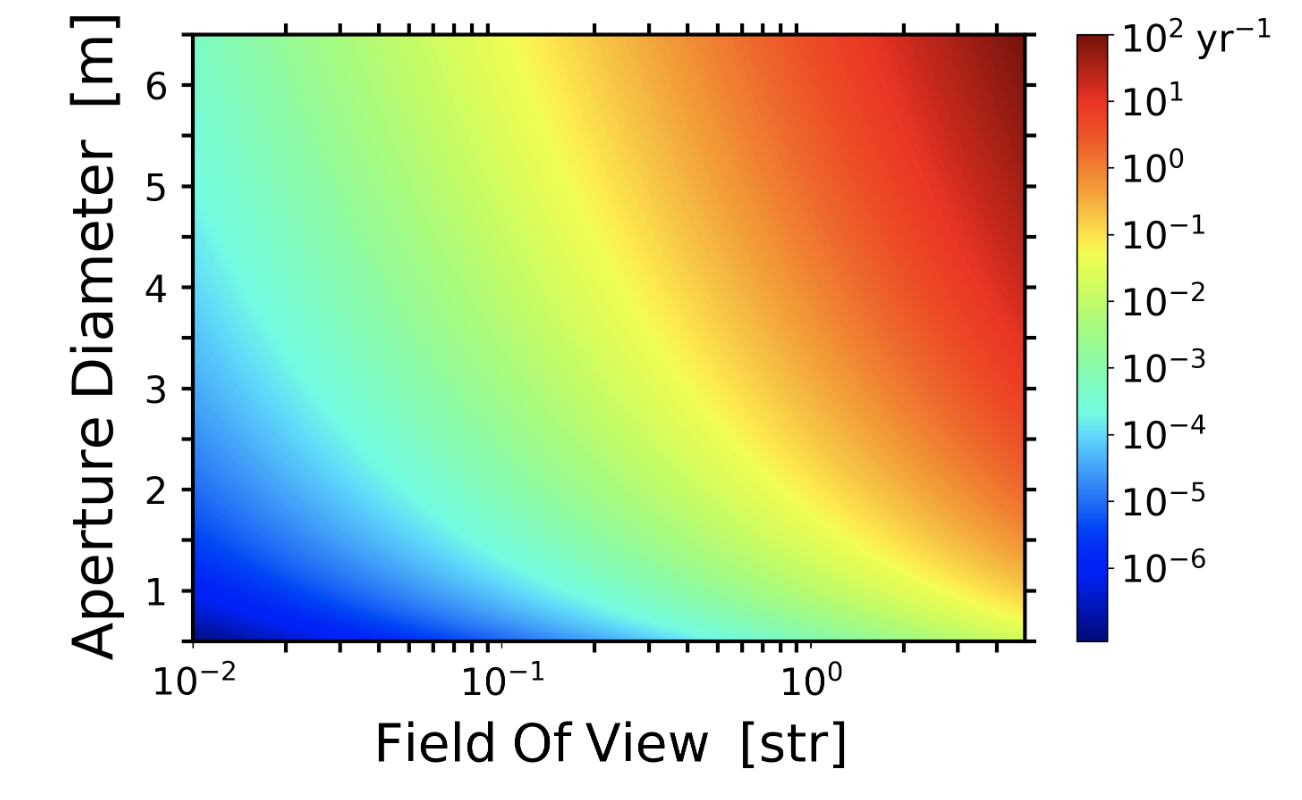}
    \caption{Expected interstellar impact rate as a function of telescope aperture diameter and field of view.}
    \label{fig:jet}
\end{figure}

\begin{figure}
  \centering
  \includegraphics[width=.8\linewidth]{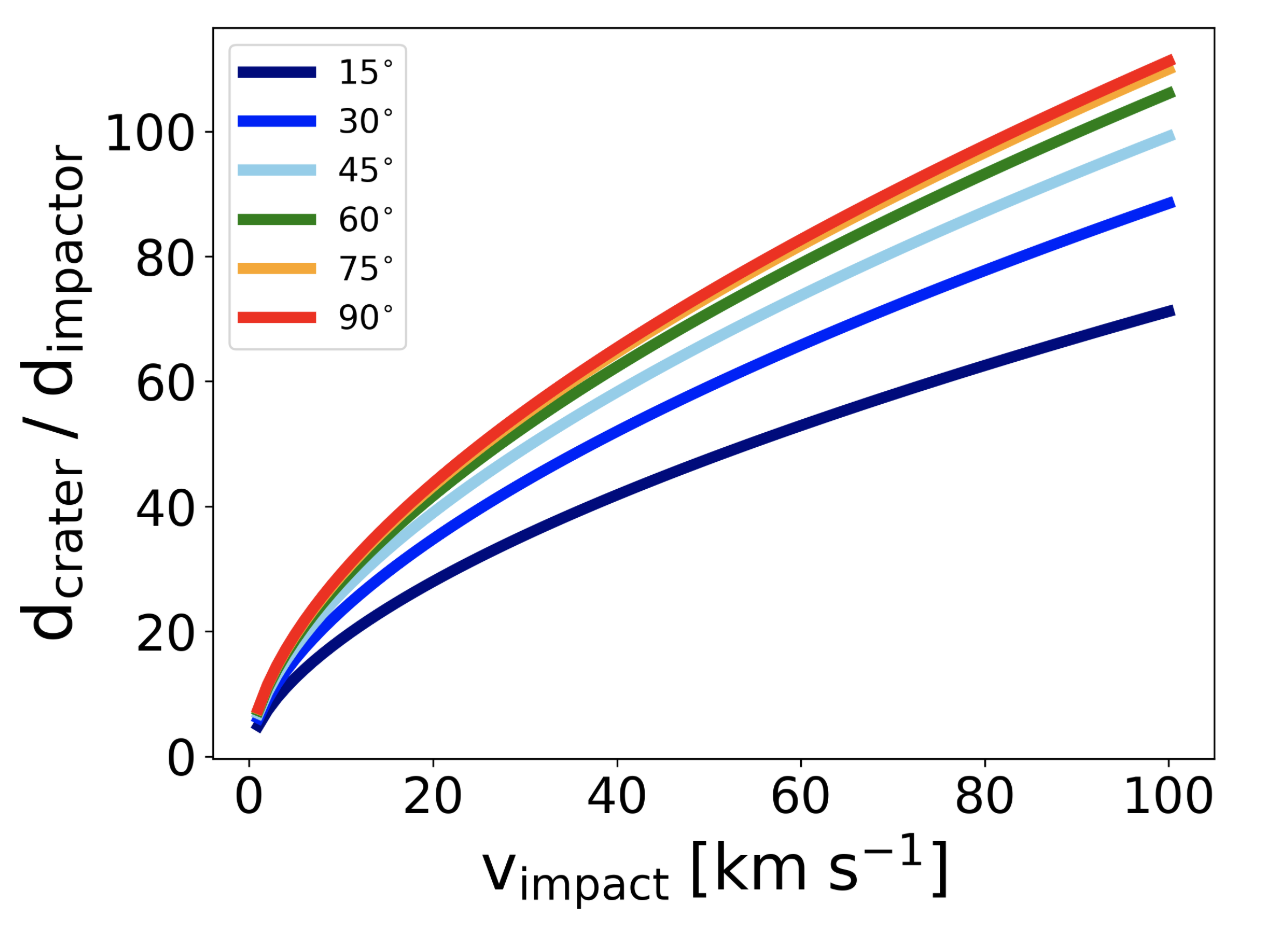}
    \caption{Relative crater size as a function of impact speed for several different impact angles.}
    \label{fig:velocities}
\end{figure}

Figure~\ref{fig:jet} shows the expected interstellar impact rate for a telescope in orbit 100 km above the lunar surface, calibrated based on CNEOS 2014-01-08 \citep{Siraj2019c,Siraj2019d}. A telescope with an aperture diameter of size $D_a \gtrsim 2 \mathrm{\; m}$ and maximal FOV ($\sim 4.9 \mathrm{\; str}$) is expected to observe $\gtrsim 1$ interstellar meteoroid impact per year. 

Figure~\ref{fig:velocities} shows the relative crater size as at various velocities and impact angles, indicating that the resulting craters should be easily detectable on the lunar surface.

\section{Discussion}
\label{sec:discussion}

We have found that a telescope in lunar orbit with diameter $D_a \gtrsim 2 \mathrm{\; m}$ would be capable of detecting $\sim 1$ interstellar meteoroid impact per year. For each meteoroid, measurements of the reflected sunlight and shadow, as well as the resulting optical flash and crater, would allow for the determination of velocity, mass, density, radiative efficiency, and composition. It is important to note that the interstellar number density calibration is based upon a single detection of CNEOS 2014-01-08, and therefore the appropriate Poisson statistics apply. The 95\% confidence bounds for an estimate of $\sim 1 \; \mathrm{yr^{-1}}$ are $0.03 - 5.57 \mathrm{\; yr^{-1}}$.

Meteoroids with known velocities and directions, such as those associated with meteor showers \citep{Rubio2000, Ortiz2015, Avdellidou2019}, can help calibrate the system.

Non-interstellar meteoroids outnumber interstellar ones by a factor of nearly $\sim 10^3$ in the centimeter-size scale \citep{Zolensky2006}, so a $D_a \sim 2 \mathrm{\; m}$ telescope that is capable of detecting a single interstellar meteoroid impact per year should detect at least one non-interstellar meteoroid per day. This provides the opportunity to study hypervelocity impacts on a daily basis, constraining the associated cratering and radiative processes \citep{Rubio2000, Drolshagen2001, Burchell2010, Ortiz2015, Ortiz2016, Goel2015, Sachse2015, Avdellidou2019}. In addition, follow-up studies of fresh craters can be performed by lunar rovers, revealing the physics of hypervelocity impacts as well as the composition of the meteoroid in more detail.

\section*{Acknowledgements}
This work was supported in part by a grant from the Breakthrough Prize Foundation. 


\begin{thebibliography}{}
\makeatletter
\relax
\def\mn@urlcharsother{\let\do\@makeother \do\$\do\&\do\#\do\^\do\_\do\%\do\~}
\def\mn@doi{\begingroup\mn@urlcharsother \@ifnextchar [ {\mn@doi@}
  {\mn@doi@[]}}
\def\mn@doi@[#1]#2{\def\@tempa{#1}\ifx\@tempa\@empty \href
  {http://dx.doi.org/#2} {doi:#2}\else \href {http://dx.doi.org/#2} {#1}\fi
  \endgroup}
\def\mn@eprint#1#2{\mn@eprint@#1:#2::\@nil}
\def\mn@eprint@arXiv#1{\href {http://arxiv.org/abs/#1} {{\tt arXiv:#1}}}
\def\mn@eprint@dblp#1{\href {http://dblp.uni-trier.de/rec/bibtex/#1.xml}
  {dblp:#1}}
\def\mn@eprint@#1:#2:#3:#4\@nil{\def\@tempa {#1}\def\@tempb {#2}\def\@tempc
  {#3}\ifx \@tempc \@empty \let \@tempc \@tempb \let \@tempb \@tempa \fi \ifx
  \@tempb \@empty \def\@tempb {arXiv}\fi \@ifundefined
  {mn@eprint@\@tempb}{\@tempb:\@tempc}{\expandafter \expandafter \csname
  mn@eprint@\@tempb\endcsname \expandafter{\@tempc}}}

\bibitem[\protect\citeauthoryear{Afanasiev, Kalenichenko \& Karachentsev}{Afanasiev et~al.}{2007}]{Afanasiev2007}
Afanasiev V. L., Kalenichenko V. V., Karachentsev I. D., 2007, \mn@doi [Astrophysical Bulletin] {10.1134/S1990341307040013}, 62(4), 319

\bibitem[\protect\citeauthoryear{Avdellidou \& Vaubaillon}{Avdellidou \& Vaubaillon}{2019}]{Avdellidou2019}
Avdellidou C., Vaubaillon J., 2019, \mn@doi [MNRAS] {10.1093/mnras/stz355}, 484, 5212

\bibitem[\protect\citeauthoryear{Baggaley et~al.}{Baggaley et~al.}{1993}]{Baggaley1993}
Baggaley W. J., Taylor, D.A. \& Steel, I.D. 1993,
Meteoroids and their Parent Bodies, Proc. Int. Astron. Symp., 53

\bibitem[\protect\citeauthoryear{Baggaley}{Baggaley}{2000}]{Baggaley2000}
Baggaley W. J., 2000, \mn@doi [Journal of Geophysical Research] {10.1029/1999JA900383}, 105(A5), 10353

\bibitem[\protect\citeauthoryear{Bannister et~al.,}{Bannister et~al.}{2017}]{Bannister2017}
Bannister M.~T.,  et~al., 2017, \mn@doi [The Astrophysical Journal]
  {10.3847/2041-8213/aaa07c}, 851, L38

\bibitem[\protect\citeauthoryear{Bialy \& Loeb}{Bialy \& Loeb}{2018}]{Bialy2018}
Bialy S.,  Loeb A.,  2018, \mn@doi [The Astrophysical Journal]
  {10.3847/2041-8213/aaeda8}, 868, L1
  
\bibitem[\protect\citeauthoryear{Bolin et~al.,}{Bolin et~al.}{2017}]{Bolin2017}
Bolin B.~T.,  et~al., 2017, \mn@doi [The Astrophysical Journal Letters, Volume
  852, Issue 1, article id. L2, 10 pp. (2018).] {10.3847/2041-8213/aaa0c9}, 852
  
\bibitem[\protect\citeauthoryear{Burchell et~al.,}{Burhcell et~al.}{2010}]{Burchell2010}
Burchell M. J., et~al., 2010, \mn@doi [Earth, Moon, and Planets] {10.1007/s11038-010-9360-5}, 107, 55

\bibitem[\protect\citeauthoryear{Charnoz \& Morbidelli}{Charnoz \& Morbidelli}{2003}]{Charnoz2003}
Charnoz S., Morbidelli A., 2003, \mn@doi [Icarus] {10.1016/S0019-1035(03)00213-6}, 166, 141

\bibitem[\protect\citeauthoryear{Do, Tucker \& Tonry}{Do et~al.}{2018}]{Do2018}
Do, A., Tucker, M. A., Tonry, J. 2018, \mn@doi [The Astrophysical Journal] {10.3847/2041-8213/aaae67}, 855, L10

\bibitem[\protect\citeauthoryear{Drolshagen}{Drolshagen}{2001}]{Drolshagen2001}
Drolshagen, G., 2001, In: Proceedings of the Meteoroids 2001 Conference, August 2001, Kiruna, Sweden. Ed.: Barbara Warmbein. ESA SP-495, Noordwijk: ESA Publications Division, pp. 533 - 541

\bibitem[\protect\citeauthoryear{Duncan, Qunn \& Tremaine}{Duncan et~al.}{1987}]{Duncan1987}
Duncan, M., Quinn, T., Tremaine, S. 1987, The Astronomical Journal, 94, 1330

\bibitem[\protect\citeauthoryear{Engelhardt et~al.,}{Engelhardt et~al.}{2017}]{Engelhardt2017}
Engelhardt T., et~al., 2017, \mn@doi [The Astronomical Journal] {10.3847/1538-3881/aa5c8a}, 153, 133

\bibitem[\protect\citeauthoryear{Fitzsimmons et~al.,}{Fitzsimmons
  et~al.}{2018}]{Fitzsimmons2018}
Fitzsimmons A.,  et~al., 2018, \mn@doi [Nature Astronomy]
  {10.1038/S41550-017-0361-4}, 2, 133
  
\bibitem[\protect\citeauthoryear{Gaidos, Williams  \& Kraus}{Gaidos
  et~al.}{2017}]{Gaidos2017}
Gaidos E.,  Williams J.,   Kraus A.,  2017, \mn@doi [Research Notes of the AAS]
  {10.3847/2515-5172/aa9851}, 1, 13
  
\bibitem[\protect\citeauthoryear{Gault}{Gault}{1974}]{Gault1974}
Gault, D. E., 1974, In: R. Greeley, P.H. Schultz (eds.), A primer in lunar
geology, NASA Ames, Moffet Field, p. 137.

\bibitem[\protect\citeauthoryear{Goel, Lee  \& Close}{Goel et~al.}{2015}]{Goel2015}
Goel A.,  Lee N.,   Close S.,  2015, \mn@doi [International Journal of Impact Engineering]
  {10.1016/j.ijimpeng.2015.05.008}, 84, 54

\bibitem[\protect\citeauthoryear{Hoang, Loeb, Lazarian \& Cho}{Hoang et~al.}{2018}]{Hoang2018}
Hoang T.,  Loeb A.,  Lazarian A.,  Cho J., 2018, \mn@doi [The Astrophysical Journal]
  {10.3847/1538-4357/aac3db}, 860(1), 42
  
\bibitem[\protect\citeauthoryear{Hajdukova}{Hajdukova}{1994}]{Hajdukova1994}
Hajdukova M., Jr., 1994, \mn@doi [Astronomy and
  Astrophysics], 288(1), 330
  
\bibitem[\protect\citeauthoryear{Hajdukova, Sterken \& Wiegert}{Hajdukova et~al.}{2018}]{Hajdukova2018}
Hajdukova M., Sterken, V., Wiegert, P., 2018, \mn@doi [European Planetary Science Congress], 12

\bibitem[\protect\citeauthoryear{Hands et~al.}{Hands et~al.}{2019}]{Hands2019}
Hands T. O., et~al., 2019, \mn@doi [MNRAS]
  {10.1093/mnras/stz1069}

\bibitem[\protect\citeauthoryear{Jewitt, Luu, Rajagopal, Kotulla, Ridgway, Liu
  \& Augusteijn}{Jewitt et~al.}{2017}]{Jewitt2017}
Jewitt D.,  Luu J.,  Rajagopal J.,  Kotulla R.,  Ridgway S.,  Liu W.,
  Augusteijn T.,  2017, \mn@doi [The Astrophysical Journal]
  {10.3847/2041-8213/aa9b2f}, 850, L36
  
\bibitem[\protect\citeauthoryear{Landgraf et~al.}{Landgraf et~al.}{2000}]{Landgraf2000}
Landgraf, M., Baggaley, W. J., Grun, E., Kruger, H., Linkert, G., 2000, J. Geophys. Res., 105, 10343

\bibitem[\protect\citeauthoryear{Lingam \& Loeb}{Lingam \& Loeb}{2019}]{Lingam2019}
Lingam M., Loeb A., 2019,  (\mn@eprint {arXiv} {1907.05427})

\bibitem[\protect\citeauthoryear{Mamajek}{Mamajek}{2017}]{Mamajek2017}
Mamajek E.,  2017, \mn@doi [Research Notes of the AAS]
  {10.3847/2515-5172/aa9bdc}, 1, 21
  
\bibitem[\protect\citeauthoryear{Mathews et~al.}{Mathews et~al.}{1998}]{Mathews1998}
Mathews, D. J., Meisel, D. D., Janches, D., Getman, S. V., Zhou, Q.-H.,  1998, Meteoroids 1998 (Proc. Int. Conf.), ed. W. J. Baggaley \& V. Porubcan (Bratislava: Astronomical Institute of the Slovak Academy of Sciences), 79

\bibitem[\protect\citeauthoryear{Meech et~al.,}{Meech et~al.}{2017}]{Meech2017}
Meech K.~J.,  et~al., 2017, \mn@doi [Nature] {10.1038/nature25020}, 552, 378

\bibitem[\protect\citeauthoryear{Meisel, Janches  \& Matthews}{Meisel et~al.}{2002a}]{Meisel2002a}
Meisel D. D.,  Janches D.,   Mathews J. D.,  2002a, \mn@doi [The Astrophysical Journal]
  {10.1086/322317}, 567, 323

\bibitem[\protect\citeauthoryear{Meisel, Janches  \& Matthews}{Meisel et~al.}{2002b}]{Meisel2002b}
Meisel D. D.,  Janches D.,   Mathews J. D.,  2002b, \mn@doi [The Astrophysical Journal]
  {10.1086/342919}, 567, 323

\bibitem[\protect\citeauthoryear{Melosh}{Melosh}{1989}]{Melosh1989}
Melosh, H. J., 1989, Impact Cratering: A Geologic Process. (Oxford Univ. Press, New York.)

\bibitem[\protect\citeauthoryear{Micheli et~al.,}{Micheli
  et~al.}{2018}]{Micheli2018}
Micheli M.,  et~al., 2018, \mn@doi [Nature] {10.1038/s41586-018-0254-4}, 559, 223

\bibitem[\protect\citeauthoryear{Moro-Martin}{Moro-Martin}{2019}]{Moro-Martin2019}
Moro-Martin A.,  et~al., 2019, \mn@doi [The Astrophysical Journal] {10.3847/1538-3881/aafda6}, 157, 86

\bibitem[\protect\citeauthoryear{Ortiz et~al.,}{Ortiz et~al.}{2015}]{Ortiz2015}
Ortiz J. L.,  et~al., 2015, \mn@doi [MNRAS] {10.1093/mnras/stv1921}, 454, 344

\bibitem[\protect\citeauthoryear{Ortiz et~al.,}{Ortiz et~al.}{2016}]{Ortiz2016}
Ortiz J. L.,  et~al., 2016, \mn@doi [Icarus] {10.1016/j.icarus.2006.05.002}, 184, 319

\bibitem[\protect\citeauthoryear{Pfalzner et~al.}{Pfalzner et~al.}{2015}]{Pfalzner2015}
Pfalzner S., et~al., 2015, \mn@doi [Phys S] {0000-0002-5003-4714}, 794, 147

\bibitem[\protect\citeauthoryear{Pfalzner \& Bannister}{Pfalzner \& Bannister}{2019}]{Pfalzner2019}
Pfalzner S., Bannister M. T., 2019, \mn@doi [The Astrophysical Journal] {10.3847/2041-8213/ab0fa0}, 874, L34

\bibitem[\protect\citeauthoryear{Musci et~al.,}{Musci et~al.}{2012}]{Musci2012}
Musci R.,  et~al., 2012, \mn@doi [The Astrophysical Journal] {10.1088/0004-637X/745/2/161}, 745, 161

\bibitem[\protect\citeauthoryear{Raymond et~al.,}{Raymond et~al.}{2018}]{Raymond2018}
Raymond S. N.,  et~al., 2018, \mn@doi [MNRAS] {10.1093/mnras/sty468}, 476, 3031

\bibitem[\protect\citeauthoryear{Rubio et~al.,}{Rubio et~al.}{2000}]{Rubio2000}
Rubio L. R. B., Ortiz J. L., Sada P. V., 2000, \mn@doi [The Astrophysical Journal] {10.1086/312914}, 542, L65

\bibitem[\protect\citeauthoryear{Sachse et~al.,}{Sachse et~al.}{2015}]{Sachse2015}
Sachse M.,  et~al., 2015, \mn@doi [Journal of Geophysical Research] {10.1002/2015JE004844}, 120(11)

\bibitem[\protect\citeauthoryear{Seligman, Luaghlin
  \& Batygin}{Seligman et~al.}{2019}]{Seligman2019}
Seligman D.,  Laughlin G.,  Batygin K., 2019, \mn@doi [The Astrophysical Journal Letters]
  {arXiv:1903.04723}, 

\bibitem[\protect\citeauthoryear{Siraj \&
  Loeb}{Siraj \&
  Loeb}{2019a}]{Siraj2019a}
Siraj A. \& Loeb A., 2019a, \mn@doi [The Astrophysical Journal]{10.3847/2041-8213/ab042a}, 872(1), L10

\bibitem[\protect\citeauthoryear{Siraj \&
  Loeb}{Siraj \&
  Loeb}{2019b}]{Siraj2019b}
Siraj A. \& Loeb A., 2019b, \mn@doi [Research Notes of the American Astronomical Society]{10.3847/2515-5172/aafe7c}, 3(1), 15

\bibitem[\protect\citeauthoryear{Siraj \&
  Loeb}{Siraj \&
  Loeb}{2019c}]{Siraj2019c}
Siraj A. \& Loeb A., 2019c, submitted to The Astrophysical Journal Letters

\bibitem[\protect\citeauthoryear{Siraj \& Loeb}{Siraj \& Loeb}{2019d}]{Siraj2019d}
Siraj A., Loeb A., 2019d,  (\mn@eprint {arXiv} {1906.03270})

\bibitem[\protect\citeauthoryear{Siraj \&
  Loeb}{Siraj \&
  Loeb}{2019e}]{Siraj2019e}
Siraj A. \& Loeb A., 2019e, submitted to MNRAS Letters

\bibitem[\protect\citeauthoryear{Suggs et~al.,}{Suggs et~al.}{2014}]{Suggs2014}
Suggs R. M., 2014, \mn@doi [Icarus] {10.1016/j.icarus.2014.04.032}, 238, 23

\bibitem[\protect\citeauthoryear{Taylor, Baggaley \& Steel}{Taylor et~al.}{1996}]{Taylor1996}
Taylor A. D., Baggaley W. J., Steel D. I., 2018, \mn@doi [Nature] {10.1038/380323a0}, 380, 323

\bibitem[\protect\citeauthoryear{Trilling et al.}{Trilling et al.}{2018}]{Trilling2018}
Trilling, D., et al.,  2018, \mn@doi [The Astronomical Journal] {	10.3847/1538-3881/aae88f},
  156, 261.

\bibitem[\protect\citeauthoryear{Veras}{Veras et~al.}{2011}]{Veras2011}
Veras D., et~al., 2011, \mn@doi [MNRAS] {10.1111/j.1365-2966.2011.19393.x}, 417, 2104

\bibitem[\protect\citeauthoryear{Veras}{Veras et~al.}{2014}]{Veras2014}
Veras D., et~al., 2014, \mn@doi [MNRAS] {10.1093/mnras/stt1905}, 437, 1127

\bibitem[\protect\citeauthoryear{Weryk \& Brown}{Weryk \& Brown}{2004}]{Weryk2004}
Weryk R. J., Brown P., 2004, \mn@doi [Icarus] {10.1007/s11038-005-9034-x},
  95, 221.
  
\bibitem[\protect\citeauthoryear{Ye, Zhang, Kelley  \& Brown}{Ye
  et~al.}{2017}]{Ye2017}
Ye Q.-Z.,  Zhang Q.,  Kelley M. S.~P.,   Brown P.~G.,  2017, \mn@doi [The
  Astrophysical Journal] {10.3847/2041-8213/aa9a34}, 851, L5
  
\bibitem[\protect\citeauthoryear{Zolensky}{Zolensky}{2006}]{Zolensky2006}
Zolensky, M., et al., 1972, Meteorites and the Early Solar System II, D. S. Lauretta and H. Y. McSween Jr. (eds.), (University of Arizona Press, Tucson), pp.869-888

\makeatother
\end{thebibliography}
\end{document}